\documentclass[12pt]{article}
\usepackage[utf8]{inputenc}
\usepackage[english]{babel}
\usepackage[margin=1in]{geometry}
\usepackage{amsmath}
\usepackage{bm}
\usepackage{amsfonts}
\usepackage{amssymb}
\usepackage{graphicx}
\usepackage{caption}
\usepackage{subcaption}
\usepackage{cite}
\usepackage{xcolor}
\usepackage{sectsty}
\sectionfont{\centering}
\usepackage{hyperref}
\hypersetup{
	colorlinks=true,
	linkcolor=blue,
	filecolor=magenta,
	urlcolor=blue,
	citecolor=violet,
	pdfpagemode=FullScreen
}
\usepackage{setspace}
\date{}
\usepackage{authblk}

\title{\large\textbf{Exclusive $J/\Psi$ and $\rho^0$ vector meson production using BK evolution theory}}

\author[1] {Ranjan Saikia 
\footnote{Corresponding author:  \href{mailto:ranjans@tezu.ernet.in}{ ranjans@tezu.ernet.in}}}

\author[1] {Jayanta Kumar Sarma \footnote{Co-author: \href{mailto:jks@tezu.ernet.in}{ jks@tezu.ernet.in}}}

\affil[1]{\small \emph{HEP Laboratory, Department of Physics, Tezpur University, Tezpur, Assam-784028, India}}
\begin{document}
\maketitle
\begin{abstract}
Exclusive diffractive processes, such as exclusive vector meson production, serve as excellent probes of hadron structure within the perturbative regime of quantum chromodynamics (QCD). The exclusive process involving light and heavy vector mesons, $e p \rightarrow e V(V=J/\Psi,\rho,\phi)$, has been investigated at the HERA accelerator facility. In this study, we focus on the theoretical prediction of exclusive $J/\Psi$ and $\rho^0$ vector meson production. Employing the color dipole description of deep inelastic scattering (DIS), we calculate both the differential cross-section and total cross-section of $J/\Psi$ and $\rho^0$ vector mesons, employing an analytical solution of the Balitsky-Kovchegov (BK) equation. Furthermore, we present the ratio of the longitudinal to the transverse cross-section for $J/\Psi$ and $\rho^0$ as a function of $Q^2$. Two well-known vector meson wave function models, Boosted Gaussian (BG) and Gaus-LC (GLC), have been integrated into our analysis, showing slight sensitivity to the chosen vector meson wave functions. Our theoretical predictions agree well with the available experimental data for vector meson production. The analytical solution of the BK equation proves reliable for the theoretical prediction of exclusive vector mesons within a specific range of $Q^2$.
  
	\vspace{1cm}
	
    {\bf Keywords:} QCD, color dipole description, vector meson production, BK equation
\end{abstract}
\thispagestyle{empty}
\clearpage
%\tableofcontents
\section{Introduction}
\label{sec:intro}
Quantum chromodynamics (QCD), also known as quantum field theory of the strong force, describes interactions between hadrons. In the limit of short distances, which corresponds to small values of the strong coupling constant $\alpha_s$, perturbative QCD (pQCD) accurately predicts the physics of strong interactions between quarks and gluons. Many deep inelastic scattering (DIS) experiments performed on hadrons by leptons at the experimental site have yielded valuable information about the distribution of partons within hadrons~[see Ref.\cite{r1} for a review]. The detailed structure of protons at the parton level has been investigated in electron-proton collisions at Hadron Electron Ring Accelerator (HERA). HERA has shown a rapid increase in gluon density at small-$x$~\cite{r2,r3}. Unfortunately, this rise cannot continue indefinitely without violating the Froissart bound and unitarity~\cite{r4}. As a result, nonlinear dynamics come to into the picture of the nuclear structure or the proton as $x$ approaches small. At small-$x$, gluon densities become so large that the smallness of $\alpha_s$ is compensated by the large gluon density. One can use pQCD to describe nonlinear dynamics at high-density QCD.

Color glass condensate (CGC) [see Refs.~\cite{r5,r6,r7} for a review] is an effective QCD theory that has been developed to deal with nonlinear dynamics in high-density QCD. It works in the proton saturation regime at small-$x$, describing the nonlinear saturation behaviour of gluons inside protons. Despite its success in describing nonlinear physics at small-$x$, we are still far from having strong evidence of the nonlinear saturation effect at present collider facilities. Some future facilities are being developed, including the Eletron-Ion Collider (EIC) in the United States~\cite{r8}, the Large Hadron electron Collider (LHeC) at CERN~\cite{r9,r10}, and the Eletron-Ion Collider in China~\cite{r11}. In these future experimental facilities, the nonlinar effects will be easily accessible as the proton will be replaced by a heavy nucleus, so that parton densities enhance nearly $A^{1/3}$ in heavy nuclei. With these advancements, future facilities are expected to address the gluon saturation phenomenon and explore other QCD aspects associated with it.

In the high-energy limit, diffractive processes such as exclusive vector meson production serve as potent tools for exploring the nonlinear dynamics of QCD in DIS experiments. This method is particularly effective for investigating the inner structure of hadrons like protons in the saturation regime at the small values of $x$~\cite{r12,r13,r14}. In such processes, there is no net transfer of color charge, which necessitates the exchange of two gluons with the target hadron at the amplitude level. Consequently, the cross-section becomes sensitive to the gluon density, and nonlinear dynamics are expected to manifest through exclusive processes. Additionally, exclusive processes like vector meson production uniquely enable the calculation of the total momentum transfer to the target, facilitating the study of generalised parton distribution functions (GPDFs)~\cite{r15,r16} and the spatial structure of target nuclei~\cite{r17,r18}.

In recent years, both theoretical and experimental efforts have focused on the exclusive photoproduction of both light and heavy vector mesons. On the experimental front, the ZEUS and H1 collaborations at the HERA accelerator facility have investigated vector mesons ($J/\Psi$, $\rho$, $\phi$)~\cite{r19,r20,r21,r22,r23,r24}, while the LHCb collaboration at CERN's LHC has provided high-precision data on exclusive $J/\Psi$ production in $pp$ collisions at energies of $\sqrt{7}$ TeV and $\sqrt{13}$ TeV~\cite{r25,r26}. On the theoretical side, the Golec-Biernat and Wusthoff (GBW) model~\cite{r27} stands as pioneering work from two decades ago, focusing on studying the gluon saturation effect using diffractive DIS processes at HERA, based on Mueller's dipole approach~\cite{r28}. Extensive research has been conducted to investigate the gluon saturation phenomenon using diffractive DIS processes. Similar to the GBW model, the Color Glass Condensate (CGC) model~\cite{r29} addresses the gluon saturation effect by describing the dipole scattering process. Two impact parameter models, the IP-Sat~\cite{r30} and b-CGC~\cite{r31} models, have been employed to assess whether exclusive vector meson production serves as a sensitive probe of gluon saturation at small-$x$. The aforementioned models are also applied to study vector meson productions in pp and nucleus-nucleus collisions at the LHC experiment~\cite{r32}. These studies demonstrated that gluon saturation models provide a qualitative description of the experimental data.

\begin{figure}[h]
	\centering
	\includegraphics[width=0.65\linewidth]{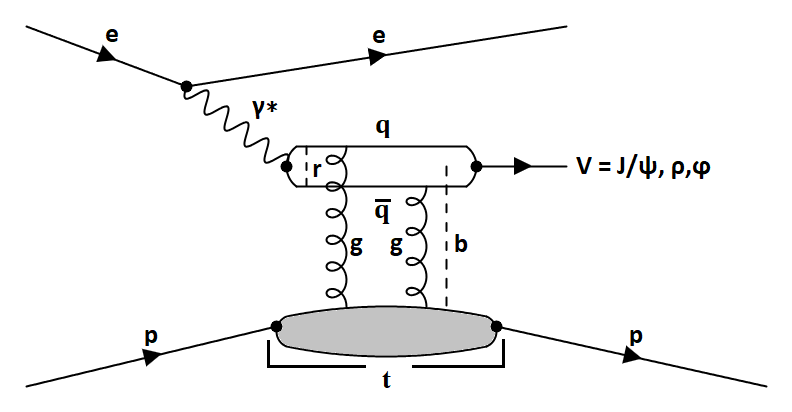}
	\caption{The exclusive vector meson production within pQCD approach in the color dipole description of e-p DIS via two gluon exchange.}
	\label{fig1}
\end{figure}

To analyze any diffractive DIS process such as vector meson production, the color dipole model of DIS serve as a powerful tool. The color dipole description of e-p DIS with two gluon exchange is depicted in Figure \ref{fig1}.
In the color dipole model, e-p DIS can be described as follows: Initially, the electron interacts with the target proton by exchanging a virtual photon. This virtual photon then splits into a quark-antiquark dipole before interacting with the proton and scattering off via gluon exchange. Finally, the dipole recombines, leading to the production of final-state particles like vector mesons or photons. The GBW model, CGC model, and CGC-based models effectively describe the dipole scattering process within this framework. In the IP-Sat and b-CGC models, the saturation effect is related to the DGLAP (Dokshitzer-Gribov-Lipatov-Altarelli-Parisi) equation~\cite{r33,r34,r35} and the BFKL (Balitsky-Kuraev-Fadin-Lipatov) equation~\cite{r36,r37,r38,r39}, respectively. The evolution of dipole-target scattering is predominantly governed by the BK (Balitsky-Kovchegov) equation~\cite{r40,r41,r42,r43}, which describes gluon saturation with a nonlinear term. The BK equation is a mean-field approximation of the more complex JIMWLK (Jalilian-Marian-Iancu-McLerran-Weigert-Leonidov-Kovner) equation~\cite{r44,r45,r46,r47}. Despite numerous numerical studies [see Refs.~\cite{r48,r49,r50,r51a,r51b}], an exact analytical solution to this equation remains elusive due to its complex nature.

Recently, we presented an approximate analytical solution to the BK equation, which proves useful for phenomenology in the context of nonlinear dynamics in high-density QCD~\cite{r52}. This solution has been successfully applied to estimate the proton structure function $F_2^ p$ within specific kinematic regions of $x$ and $Q^2$ at HERA~\cite{r53}. In Ref.~\cite{r52,r54}, the exclusive vector meson production has been studied using a numerical solution to the BK equation. In this work, we utilise the analytical solution of the BK equation to investigate the exclusive vector meson production process and its dependence on various scales. Within the color dipole description of DIS, we estimate the theoretical measurements of exclusive vector meson ($J/\Psi, \rho^{0}$) production, including: $(i)$ the differential cross-section with the four-momentum transfer squared at the proton vertex, $t$, and $(ii)$ the total cross-section as functions of the center-of-mass energy of the photon-photon system, $W$, and the exchanged-photon virtuality, $Q^2$. The novel aspect of the present analysis is that one of the ingredients for the measurement of elxclusive vector meson production within the color dipole description is obtained from the analytical solution of the BK equation.

The structure of the paper is as follows: In Section \ref{sec2}, we discuss and review the color dipole description of exclusive vector meson production in e-p DIS. The approximate analytical solution of the BK equation is introduced in Section \ref{sec3}. In Section \ref{sec4}, we present the theoretical estimation of the differential and total cross-sections for the production of $J/\Psi$ and $\rho^0$ vector mesons in the process $e p \rightarrow e p V$ ($V = J/\Psi, \rho^0$), considering the various scales involved. Additionally, the ratio of the longitudinal to the transverse cross-section for $J/\Psi$ and $\rho^0$, as a function of $Q^2$, is presented in this section. We compare our theoretical results with existing experimental data on vector meson production. Section \ref{sec5} provides a summary of our work, followed by conclusions.
\section{Exclusive vector meson production within color dipole description of DIS}\label{sec2}
In this section, we briefly review and discuss the dipole description of DIS for the production of exclusive vector mesons. We first provide the dipole scattering amplitude for exclusive vector mesons within the CGC framework. We then discuss wavefunctions for vector mesons, which are a key portion of the dipole description of DIS for vector meson production.
\subsection{Dipole scattering amplitude for vector mesons}
In the color dipole description of DIS, a simple e-p DIS ($e p \rightarrow e p V$ ($V = J/\Psi, \rho, \phi$)) is considered as the interaction of the color dipole (quark-antiquark pair) with the target proton. The dipole scattering amplitude for exclusive vector mesons consists of three ingredients: the photon wave function, the dipole-proton scattering amplitude, and the vector meson wave function. A schematic diagram for the exclusive vector meson production within the pQCD approach in the color dipole description of e-p DIS via two gluon exchange is depicted in Figure \ref{fig1}. The transverse size of the quark-antiquak pair is expressed as $\mathbf{r}$, and $\mathbf{b}$ is the impact parameter for the dipole-proton interaction. The quark carries a fraction of monemtum ($z$) of the photon's light-cone momentum and ($1-z$) for the antiquark. In the perturbative region, which corresponds to the smaller dipole sizes ($r$), the dipole-proton cross section corresponds to the exchange of a gluon ladder. In the rest frame of the proton, one can express the imaginary part of the scattering amplitude for vector meson production as~\cite{r55}
\begin{equation}\label{eqn1}
	\mathcal{A}^{\gamma^{\ast} p \rightarrow V p} _{T, L} (x, Q^2 , \mathbf{\Delta}) = i \int d^2 \mathbf{r} \int _0 ^1 \frac{dz}{4\pi} \int d^2 \mathbf{b} (\Psi ^{\ast} _V \Psi)_{T,L} e^{-i\mathbf{b}.\mathbf{\Delta}} \times 2[1-S(x,r,b)].
\end{equation}
Here $x$ represents the Bjorken variable, while $Q^2$ denotes the virtuality of the photon. The symbol $\mathbf{\Delta}$ represents the transverse momentum lost at the proton vertex by the outgoing proton, which is correlated with the squared momentum transfer as $t= - \mathbf{\Delta}^2$. $\Psi$ denotes the wave function for the incoming photon in the process, a well-established concept in Quantum Electrodynamics (QED). $\Psi^{\ast} _V$ is the wave function for the final state vector meson. $(\Psi ^{\ast}_V \Psi)_{T,L}$ represents the overlap of the wave functions between the photon and the vector meson, where $T$ signifies transversely polarized particles and $L$ signifies longitudinally polarized particles. This expression quantifies the probability amplitude for the virtual photon to produce an exclusive vector meson or a real photon in the final state in e-p DIS within the framework of the color dipole description.

The scattering amplitude shown in Eq. \eqref{eqn1} contains only the forward component of the amplitude. To have the non-forward scattering amplitude, we multiply the usual forward wave functions by an exponential factor, exp$[\pm i(1-z)\mathbf{r}. \mathbf{\Delta}/2]$, following the work done by Bartels et. al. in Ref.~\cite{r56}. Following this change, together with the assumption that the S-matrix is purely real, we may write the scattering amplitude Eq. \eqref{eqn1} as 
\begin{equation}\label{eqn2}
		\mathcal{A}^{\gamma^{\ast} p \rightarrow V p} _{T, L} (x, Q^2 , \mathbf{\Delta}) = i \int d^2 \mathbf{r} \int _0 ^1 \frac{dz}{4\pi} \int d^2 \mathbf{b} (\Psi ^{\ast} _V \Psi)_{T,L} e^{-i[ \mathbf{b} -(1-z)\mathbf{r}] .\mathbf{\Delta}} \frac{d\sigma_{q\bar{q}}}{d^2 \mathbf{b}},
\end{equation}
where $\frac{d\sigma_{q\bar{q}}}{d^2 \mathbf{b}} = 2[1-\text{Re} \hspace{1mm} S(x,r,b)]$ is the dipole-proton differential scattering cross section, with $\mathbf{b}$ being the impact parameter. Here $\sigma_{q\bar{q}}$ is the dipole-proton scattering cross section, which can be related to the forward scattering amplitude, $N(x,r,b)$ for the dipole-proton system. The scattering amplitude will come from the solution of the dipole evolution equation, in our case, the BK equation. 

The differental cross section for the final state exclusive vector meson is given by~\cite{r55}
	\begin{equation}\label{eqn3}
		\frac{d\sigma^{\gamma^{\ast} p \rightarrow V p} _{T, L}}{dt} = \frac{R_g ^2}{16\pi}\big|\mathcal{A}^{\gamma^{\ast} p \rightarrow V p} _{T, L}\big|^2 \big(1+\beta^2\big),
	\end{equation}
where $\beta$ is the ratio of real to imaginary parts of the scattering amplitude used to determine the real part of the scattering amplitude, given by
\begin{equation}\label{eqn4}
	\beta=\tan(\pi\lambda/2), \hspace{5mm} \lambda = \frac{\partial\ln\big(\mathcal{A}^{\gamma^{\ast} p \rightarrow V p} _{T, L}\big)}{\partial\ln(1/x)}.
\end{equation}
The skewed effect is reflected by $R_g ^2$, given by~\cite{r57}
\begin{equation}\label{eqn5}
	R_g = \frac{2^{2\lambda+3}}{\sqrt{\pi}}\frac{\Gamma(\lambda+5/2)}{\Gamma(\lambda+4)}.
\end{equation}
For the exponential dependence of $t$ on the scattering amplitude $\mathcal{A}^{\gamma^{\ast} p \rightarrow V p} _{T, L}$~\cite{r58}, one can rewrite Eq. \eqref{eqn3} as follows:
\begin{equation}\label{eqn6}
	\frac{d\sigma^{\gamma^{\ast} p \rightarrow V p} _{T, L}}{dt} (x, Q^2 , t) = \frac{R_g ^2}{16\pi}\big|\mathcal{A}^{\gamma^{\ast} p \rightarrow V p} _{T, L}\big|_{t=0} ^2 \big(1+\beta^2\big) e^{-B_D |t|},
\end{equation}
where $B_D$ denotes the area size of the interaction region, which can be obtained with a fit to the $t$-ditributions of the form $d\sigma/dt \propto \text{exp} (-B_D|t|)$. The total scattering cross section is then obtained as 
\begin{equation}\label{eqn7}
	\sigma^{\gamma^{\ast} p \rightarrow V p} _{tot} (x, Q^2) = \sigma^{\gamma^{\ast} p \rightarrow V p} _{T} (x, Q^2) + \sigma^{\gamma^{\ast} p \rightarrow V p} _{L} (x, Q^2),
\end{equation}
with 
\begin{equation}\label{eqn8}
	\sigma^{\gamma^{\ast} p \rightarrow V p} _{T,L} (x, Q^2) = \frac{R_g ^2}{16\pi B_D}\big|\mathcal{A}^{\gamma^{\ast} p \rightarrow V p} _{T, L}\big|_{t=0} ^2 \big(1+\beta^2\big).
\end{equation}
For the $\rho^0$ vector meson, $B_D$ is given by~\cite{r58}
\begin{equation}\label{eqn9}
	B_D = N\Big(14.0\Big(\frac{1 GeV^2}{Q^2 + M_V ^2}\Big)^{0.2} +1\Big),
\end{equation}
with $N = 0.55 \hspace{1mm} GeV^{-2}$ and $M_V = 0.776 \hspace{1mm} GeV$ for $\rho^0$ meson. For the $J/\Psi$ vector meson, $B_D$ is given by~\cite{r59}
\begin{equation}\label{eqn10}
	B_D = \begin{cases}
		4.15 + 4 \times 0.116 \ln \big(\frac{W}{W_0}\big), & Q^2 \le 1 \hspace{1mm} GeV^2 \\
		4.72 + 4 \times 0.07 \ln \big(\frac{W}{W_0}\big), &  Q^2 > 1 \hspace{1mm} GeV^2 
	\end{cases},
\end{equation}
where $W_0$ $= 90 \hspace{1mm}GeV$ and $W$ is center of mass energy for $\gamma^{\ast} p$ system, related to $x$ and $Q^2$ by
\begin{equation}\label{eqn11}
	x = x_{Bj} \big(1+ \frac{M_V ^2}{Q^2}\big) = \frac{Q^2 +M_V ^2}{W^2 + Q^2},
\end{equation}
where $x_{Bj}$ is the Bjorken scale and $M_V = 3.097 \hspace{1mm}GeV$ for $J/\Psi$ vector meson.
\subsection{Vector meson wavefunctions}
One of the key ingredients to performing the production of exclusive vector mesons is the overlap wave function $(\Psi ^{\ast} _V \Psi)_{T,L}$, which is a function of the longitudinal momentum fraction $z$ carried by the quark, the dipole size $r$, and the virtuality of the photon, $Q^2$. In the literature, there are many different prescriptions for the overlap wavefunction between the photon and the vector meson, such as the DGKP (Dosch-Gousset-Kulzinger-Pirner) model~\cite{r60}, the Gaus-LC (GLC)~\cite{r61}, and the boosted Gaussian (BG) model~\cite{r62}, first proposed by Nemchik et. al~\cite{r62a,r62b}. The overlap function between the photon and the vector meson wave functions can be written as 
\begin{equation}\label{eqn12}
	(\Psi ^{\ast} _V \Psi)_{T} = \hat{e}_f e \frac{N_c}{\pi z (1-z)} \{m_f ^2 K_0 (\epsilon r)\phi_T (r,z) - [z^2 + (1-z)^2] \epsilon K_1 (\epsilon r) \partial _r \phi_T (r,z)\},
\end{equation}
\begin{equation}\label{eqn13}
	(\Psi ^{\ast} _V \Psi)_{L} = \hat{e}_f e \frac{N_c}{\pi} 2Qz (1-z) K_0 (\epsilon r) \Big[M_V \phi_L (r,z) + \delta\frac{m_f ^2 -\Delta_r ^2}{M_V z(1-z)}\phi_L(r,z)\Big],
\end{equation}
where $\hat{e}_f$ is the effective charge ($\hat{e}_f = 2/3, 1/\sqrt{2}$ for $J/\Psi$ and $\rho$ mesons respectively.), $e=\sqrt{4\pi \alpha_{em}}$, $N_c$ ($=3$) is the number of colors, $\Delta_r ^2 = (1/r)\partial_r + \partial _r ^2$, $\epsilon = \sqrt{z(1-z)Q^2 + m_f ^2}$ ($m_f$ is the quark mass) and $K_0$, $K_1$ are the second kind Bessel function. $\phi_T$ and $\phi_L$ are the scalar parts of the overlap wave functions. 

The vector meson wave functions are constrained by model-independent features. Firstly, it should satisfy the following normalization condition: 
\begin{equation}\label{new1}
	1=\sum_{h, \bar{h}}\int d^2 \mathbf{r} \int_{0}^{1} \frac{dz}{4\pi} {\Big|\Psi^V _{h,\bar{h}} (z, \mathbf{r})\Big|}^2 .
\end{equation}
Here, $\Psi^V _{h,\bar{h}} (z, \mathbf{r})$ represents the light-cone wave function of the vector meson with quark and antiquark helicities labelled by $h$ and $\bar{h}$, respectively. The above normalization condition neglected the possible contributions of gluon or sea-quark states to the vector meson wave function, assuming that the quantum numbers of the meson are saturated by the quark-antiquark pair. For the scalar parts of the vector meson wave functions, the normalization conditions are~\cite{r62a,r62b}
\begin{equation}\label{new2}
	1 = \frac{N_c}{2\pi}\int_{0}^{1}\frac{dz}{z^2 (1-z)^2} \int d^2 \mathbf{r} \hspace{2mm} \{m_f ^2 \phi_T ^2 + [z^2 + (1-z)^2]\hspace{1.5mm}(\partial_r \phi_T)^2\},
\end{equation}
\begin{equation}\label{new3}
	1 = \frac{N_c}{2\pi}\int_{0}^{1} dz \int d^2 \mathbf{r} \hspace{2mm} \Big[M_V \phi_L + \delta \frac{m_f ^2 - \Delta_r ^2}{M_V z (1-z)} \phi_L\Big]^2 .
\end{equation}
In addition to that, another important constraint comes from the leptonic decay width $\Gamma (V \rightarrow e^{+}e^{-})$, given by
\begin{equation}\label{new4}
	f_{V, T} = \hat{e}_f \frac{N_c}{2\pi M_V} \int_{0}^{1} \frac{dz}{z^2 (1-z)^2} \hspace{2mm}\{m_f ^2 - [z^2 + (1-z)^2] \Delta_r ^2\} \hspace{1.5mm}\phi_T (r, z) \Big| _{r=0} \hspace{2mm} ,
\end{equation}
\begin{equation}\label{new5}
	f_{V, L} = \hat{e}_f \frac{N_c}{\pi} \int_{0}^{1} dz \hspace{2mm}\Big[M_V + \delta \frac{m_f ^2 - \Delta_r ^2}{M_V z(1-z)} \Big] \hspace{1.5mm}\phi_L(r,z) \Big|_{r=0} \hspace{2mm},
\end{equation}
where $f_V$ is the coupling of the meson to the electromagnetic current obtained from the measured electronic decay width by
\begin{equation}\label{new6}
	\Gamma_{V\rightarrow e^{+}e^{-}} = \frac{4\pi \alpha_{em} ^2 f_V ^2}{3 M_V}.
\end{equation}

For the completeness of the vector meson wave functions [\eqref{eqn12}, \eqref{eqn13}], the scalar parts ($\phi_{T,L}$) of the vector meson wave functions should be mentioned. In this work, we employ the BG and GLC models for the scalar part wave functions as they provide a better description of existing experimental data using the constraints given in Eqs. \eqref{new2}, \eqref{new3}, \eqref{new4}, and \eqref{new5}. In the literature, there are a lot of other models that provide predictions for electroproduction cross sections of various heavy quarkonium states with good description of experimental data (see Ref.~\cite{r59}, for example). We set $\delta=0$ in the GLC model and $\delta = 1$ in the BG model, following the work done by Kowalski et. al.~\cite{r55}. 

The scalar-part wave functions in the GLC model are given by
\begin{equation}\label{eqn14}
	\begin{aligned}
		& \phi_T (r,z)= N_T [z(1-z)]^2 \hspace{1mm}\text{exp}(-r^2 /2R_T ^2),\\
		& \phi_L (r,z) = N_L z(1-z)\hspace{1mm} \text{exp}(-r^2 /2R_L ^2).
	\end{aligned}
\end{equation}
The parameters of the GLC model are given in Table \ref{tab1}.
\begin{table}[h!]
	\centering
	\begin{tabular}{ c c c c c c c c }
		\hline 
		
		Meson & $M_V/GeV$ & $f_V$ & $m_f/GeV$ & $N_T$ & $R_T ^2/GeV^{-2}$ & $N_L$ & $R_L ^2/GeV^{-2}$\\
		 \hline
		$J/\Psi$ & $3.097$ & $0.274$ &  $1.4$ & $1.23$ & $6.5$ & $0.83$ & $3.0$\\  
		$\rho^0$ & $0.776$ &$0.156$ & $0.14$ & $4.47$ & $21.9$ & $1.79$ & $10.4$\\
		\hline
	\end{tabular}
	\caption{{\label{tab1}}Parameters of the GLC model for $J/\Psi$ and $\rho^0$ vector mesons~\cite{r61}.}
\end{table}

\begin{figure}[h]
	\centering
	\begin{subfigure}{0.45\textwidth}
		\centering
		\includegraphics[width=\textwidth]{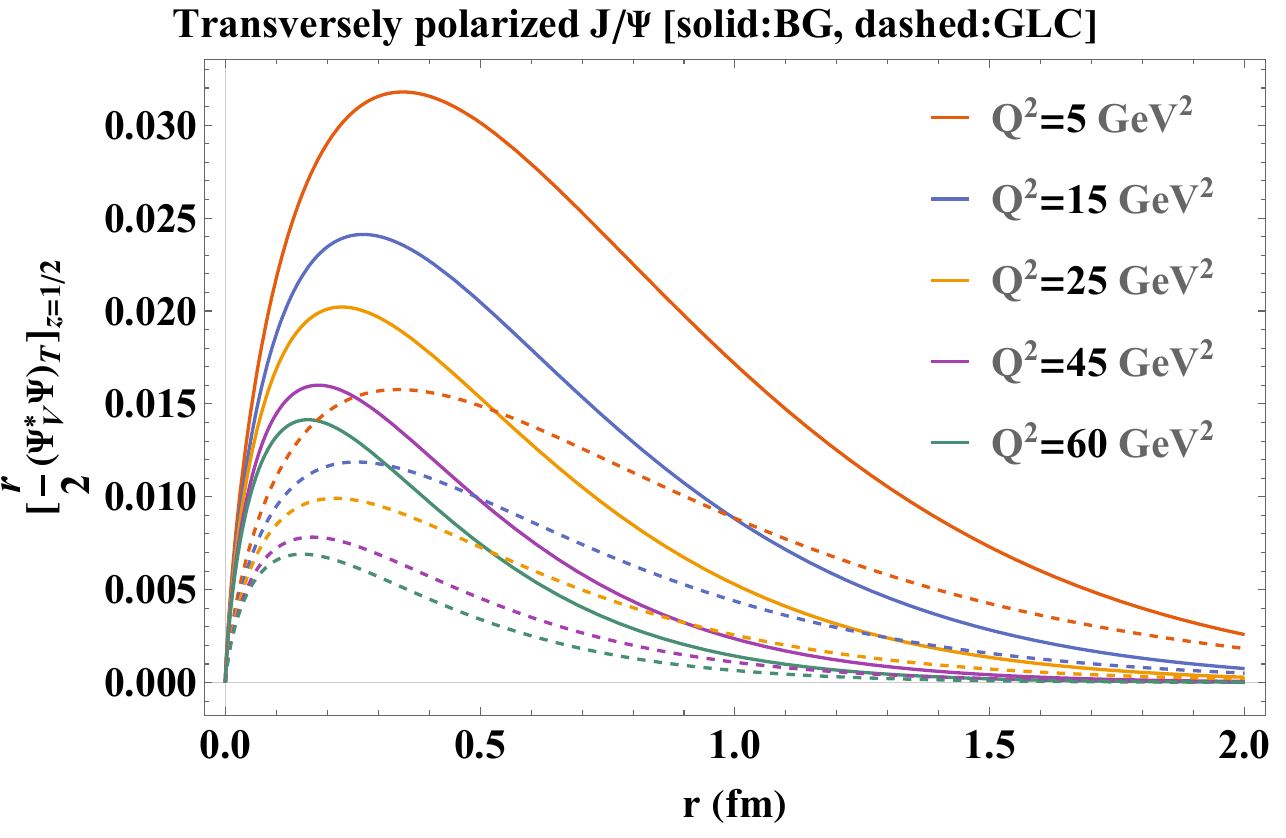}
	\end{subfigure}
	\hfill
	\begin{subfigure}{0.45\textwidth}
		\centering
		\includegraphics[width=\textwidth]{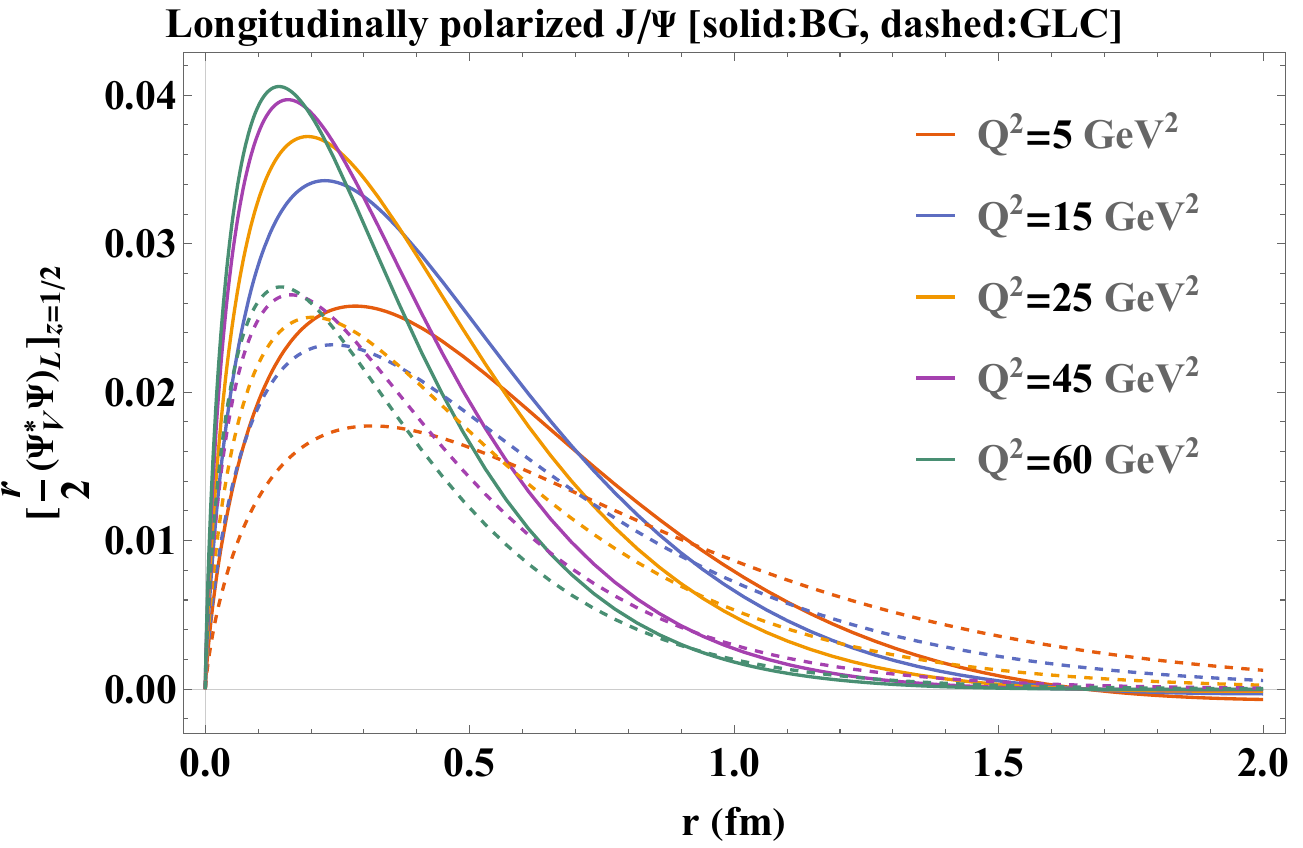}
	\end{subfigure}
	
	\begin{subfigure}{0.45\textwidth}
		\centering
		\includegraphics[width=\textwidth]{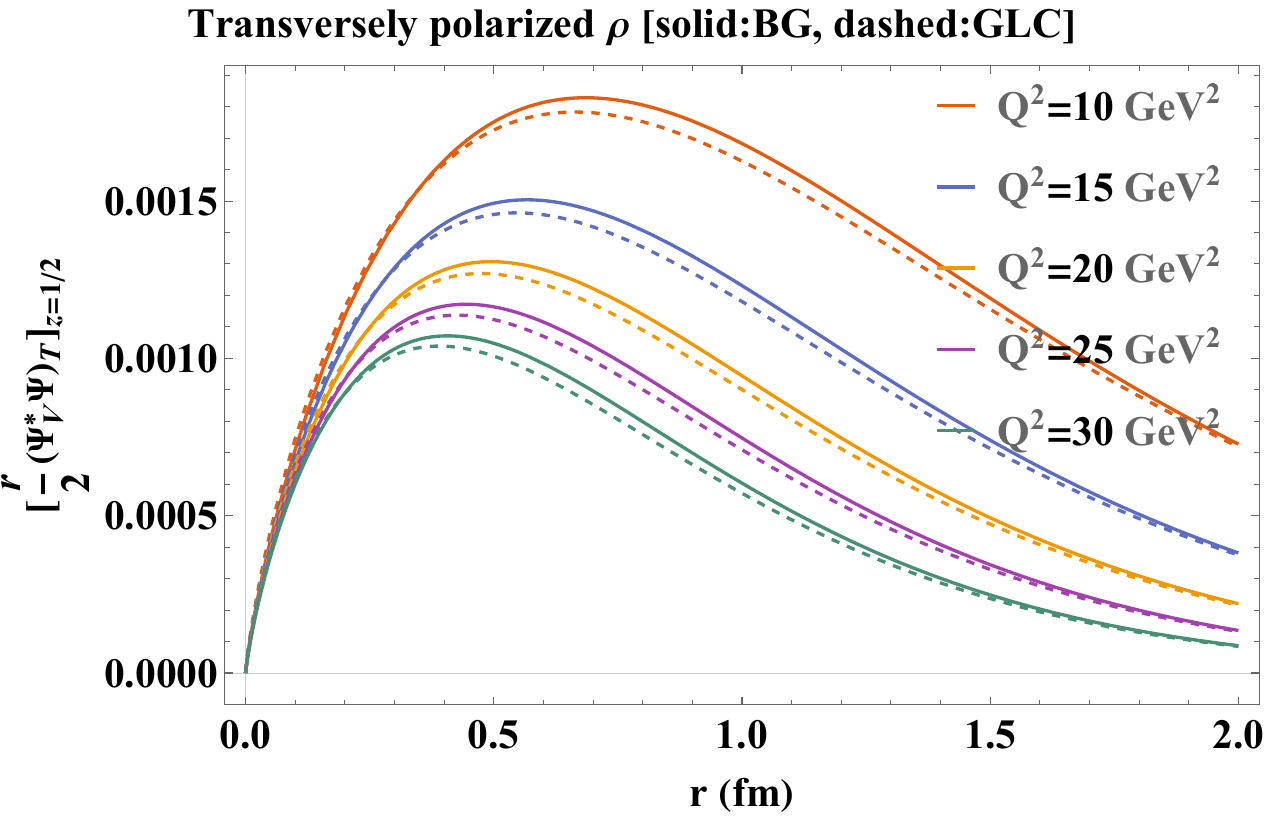}
	\end{subfigure}
	\hfill
	\begin{subfigure}{0.45\textwidth}
		\centering
		\includegraphics[width=\textwidth]{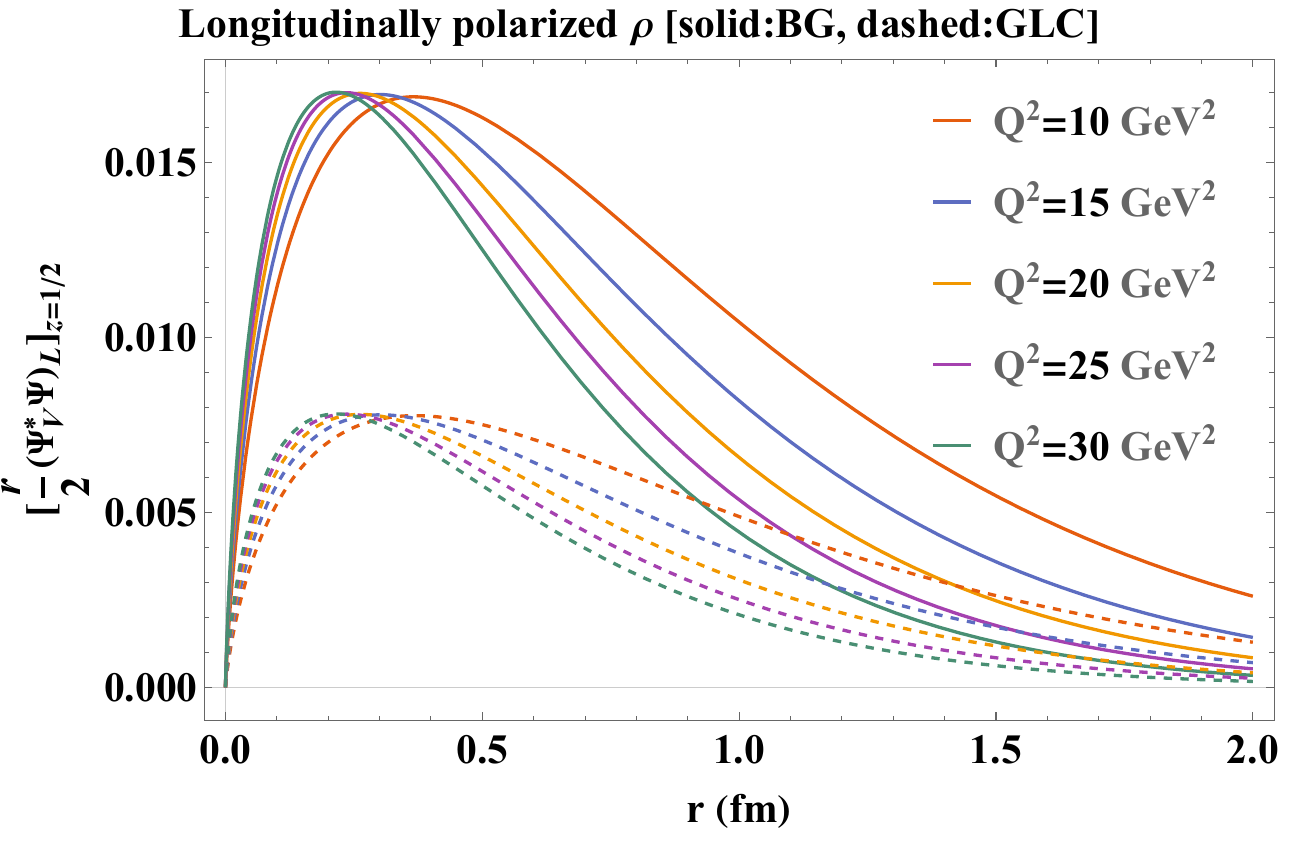}
	\end{subfigure}
	\caption{{\label{fig2}}Overlap function between the photon and the vector meson wave functions [Eqs. \eqref{eqn12} and \eqref{eqn13}], as a function of dipole size $r$ at $z=1/2$ for various $Q^2$ values.}
\end{figure}

The scalar-part wave functions in the BG model are given by
\begin{equation}\label{eqn15}
	\begin{aligned}
		& \phi_T (r,z) = \mathcal{N_T} z(1-z)\hspace{1mm} \text{exp}\Big(-\frac{m_f ^2 \mathcal{R}^2}{8z(1-z)}-\frac{2z(1-z)r^2}{\mathcal{R}^2}+ \frac{m_f ^2 \mathcal{R}_T ^2}{2}\Big),\\
		& \phi_L (r,z) = \mathcal{N_L} z(1-z)\hspace{1mm} \text{exp}\Big(-\frac{m_f ^2 \mathcal{R}^2}{8z(1-z)}-\frac{2z(1-z)r^2}{\mathcal{R}^2}+ \frac{m_f ^2 \mathcal{R}_L ^2}{2}\Big).
	\end{aligned}
\end{equation}
The parameters of the BG model are given in Table \ref{tab2}.
\begin{table}[h!]
	\centering
	\begin{tabular}{ c c c c c c c c }
		\hline 
		
		Meson & $M_V/GeV$ & $f_V$ & $m_f/GeV$ & $\mathcal{N_T}$  & $\mathcal{N_L}$ & $\mathcal{R_T} ^2/GeV^{-2}$ & $\mathcal{R_L} ^2/GeV^{-2}$\\ 
		\hline
		$J/\Psi$ & $3.097$ & $0.274$ &  $1.4$ & $0.578$ & $0.575$ & $2.3$ & $2.3$\\  
		$\rho^0$ & $0.776$ & $0.156$ & $0.14$ & $0.911$ & $0.853$ & $12.9$ & $12.9$ \\
		\hline 
	\end{tabular}
	\caption{{\label{tab2}}Parameters of the BG model for $J/\Psi$ and $\rho^0$ vector mesons~\cite{r62}.}
\end{table}
We incorporate these two models in our calculations for vector meson production, which are slightly sensitive to the given vector meson wave functions, and compare the effects of the two models on the results. For the sake of visualisation, we present the transverse and longitudinal overlaps between the vector meson and the photon wave functions as a function of the dipole size $r$ at $z=1/2$ $(0<z<1)$ in Figure \ref{fig2}. 
\section{Analytical solution to the BK equation}\label{sec3}
Another key ingredient in the calculation of exclusive vector meson processes is the dipole-target scattering amplitude. The scattering amplitude describes the propagation of the quark-antiquark dipole through the target color field. The dipole-proton cross section, $\sigma_{q\bar{q}}(x,r)$, is written as
\begin{equation}\label{eqn16}
	\sigma_{q\bar{q}}(x,r) = \int\frac{d\sigma_{q\bar{q}}}{d^2 \mathbf{b}} (x,r,\mathbf{b}) d^2 \mathbf{b} = 2\pi R_p ^2 N(x,r),
\end{equation}
where $R_p$ is the proton radius, and for this work, its value will be taken from recent work on the radius of the proton~\cite{r63}, given by $R_p\approx 0.831 \hspace{1mm} \text{fm} \approx 4.22 \hspace{1mm}  GeV^{-1}$. The scattering amplitude $N(x,r)$ comes from the BK evolution equation, which describes the evolution of the scattering amplitude $N(x,r)$ in the color diple description.

The BK equation in momentum space reads~\cite{r42}
\begin{equation}\label{eqn17}
	\partial_Y N (k,Y) = \bar{{\alpha}}\chi(-\partial_L)N (k,Y) - \bar{{\alpha}}N^2 (k,Y) , 
\end{equation}
where $N (k,Y)$ is the scattering amplitude in momentum space with transverse momentum $k$ and total rapidity $Y$ ($Y=\ln 1/x$), $\bar{{\alpha}}=\frac{\alpha_s N_c}{\pi}$, and $\chi (\gamma) = 2\psi(1) -\psi(\gamma) -\psi(1-\gamma)$ is the BFKL kernel ($\gamma = -\partial_L$, $L =\ln k^2 / k_0 ^2$). In the work done by Munier et al.~\cite{r64,r65}, it has been shown that the expansion of the BFKL kernel around $\gamma =1/2$ the BK Eq. \eqref{eqn17} reduces to the following nonlinear equation given by
\begin{equation}\label{eqn18}
	    \partial_Y N (k,Y) = \bar{{\alpha}}\bar{\chi}(-\partial_L)N (k,Y) - \bar{{\alpha}}N^2 (k,Y) , 
\end{equation}
with $\bar{\chi}(-\partial_L) = \chi(1/2) + \frac{\chi^{\prime\prime}(1/2)}{2}(\partial_L + 1/2)^2$. With this expansion, together with some changes in variables~\cite{r64}, Eq. \eqref{eqn18} can be reexpressed as the FKPP (Fisher-Kolmogorov-Petrovsky-Piscounov) equation~\cite{r66,r67} given by
\begin{equation}\label{eqn19}
	\partial_t u (t,x) = \partial_x ^2 u (t,x) + u (t,x) - u^2 (t,x) ,
\end{equation}
where
\begin{equation}\label{eqn20}
	u (t,x) = \frac{2}{\chi^{\prime\prime}(1/2) (1-\bar{\xi})^2} \times N \Big(\frac{2t}{\bar{{\alpha}}\chi^{\prime\prime}(1/2) (1-\bar{\xi})^2} , \frac{x}{1-\bar{\xi}} - \frac{t}{(1-\bar{\xi})^2}\Big),
\end{equation}
with $t = \frac{\bar{\alpha}\chi^{\prime\prime} (1/2)}{2} (1-\bar{\xi})^2 Y$, $x = (1-\bar{\xi}) \Big(L + \frac{\bar{\alpha}\chi^{\prime\prime}(1/2)}{2} Y\Big)$,
and $\bar{\xi}= 1 - \frac{1}{2}\sqrt{1+ 8 \frac{\chi(1/2)}{\chi^{\prime\prime} (1/2)}}$.
In Ref.~\cite{r52}, we present an approximate analytical solution to the BK Eq. \eqref{eqn18} in connection with the FKPP Eq. \eqref{eqn19} given by 
\begin{equation}\label{eqn21}
    N(k,Y) = \frac{Ne^Y}{1-N + N e^Y},
\end{equation} 
where $N$ is the initial condition at $Y=0$, i.e., $N(k, Y=0) = N$. The solution Eq. \eqref{eqn21} provides the dipole-proton scattering amplitude at higher rapidity $Y>0$, provided the initial condition $N$. This scattering amplitude contains all the information regarding the strong interaction at small-$x$ within the color dipole description. In our solution, we use the GBW (Golec-Biernat and Wusthoff) initial condition for $N$, given by~\cite{r27}
\begin{equation}\label{eqn22}
	N(r, Y=0) = 1 - \text{exp} \Big[-\Big(\frac{r^2 Q_{s0} ^2}{4}\Big)\Big],
\end{equation}
where $Q_{s0} ^2$ denotes the squared initial saturation momentum for gluons that can be fitted from the existing HERA data. The value of $Q_{s0} ^2$ is $0.24$ $GeV^2$~\cite{r68}. To use the GBW initial condition in our solution, let us transform the coordinate space GBW initial condition Eq. \eqref{eqn22} into the momentum space using the following simple Fourier transform, given by
\begin{equation}\label{eqn23}
	N(k, Y=0) \int \frac{d^2 r}{2\pi r^2} e^{ik.r} N(r,Y=0) = \frac{1}{2}\Gamma \Big(0, \frac{k^2}{Q_{s0}^2}\Big),
\end{equation}
resulting in the incomplete gamma function $\Gamma$. We can express the incomplete gamma function at large values of $k^2 / Q_{s0}^2$ as
\begin{equation}\label{eqn24}
	\Gamma \Big(0, \frac{k^2}{Q_{s0}^2}\Big) = \text{exp} \Big(- \frac{k^2}{Q_{s0}^2}\Big).
\end{equation}
With this, we can express the GBW initial condition Eq. \eqref{eqn22} in momentum space as 
\begin{equation}\label{eqn25}
    N(k,Y=0) = \frac{1}{2} \text{exp} \Big(-\frac{k^2}{Q_{s0}^2}\Big).
\end{equation}
Now, if we replace the initial condition $N$ in BK solution Eq. \eqref{eqn21} with the above momentum space GBW initial condition, we obtain
\begin{equation}\label{eqn26}
	N(k,Y) = \frac{e^{Y-k^2/Q_{s0}^2}}{1-e^{-k^2/Q_{s0}^2} + e^{Y-k^2/Q_{s0}^2}}.
\end{equation}
This is an approximate analytical solution to the BK equation in momentum space. In our calculations, we will need the dipole scattering amplitude in coordinate space. The dipole sacttering amplitude in coordinate space is related to the momentum space by the following Fourier transformation, given by
\begin{equation}\label{eqn27}
	N(k,Y) = \int\frac{d^2 r}{2\pi r^2} e^{i k.r} N(x,r).
\end{equation} 
With the inverse Fourier transformation, we obtain $N$ in coordinate space, given by
\begin{equation}\label{eqn28}
	N(x,r) = \frac{r^2}{2\pi}\int d^2 k e^{-ik.r} N(k,Y) = r^2\int_0 ^ \infty dk k J_0 (k.r) N(k,Y).
\end{equation}
We use Eq. \eqref{eqn28} together with vector meson wavefunctions discussed in the calculation of the differential and the total cross sections of the $J/\Psi$ and the $\rho^0$ vector mesons.
\section{Results and discussion}\label{sec4}
In this section, we present the numerical results for the cross sections of the $J/\Psi$ and $\rho^0$ vector mesons as a function of various scales involved in the process $\gamma^{\ast} p \rightarrow V p$. We present the theoretical estimation of the differential cross sections and the total cross sections of $J/\Psi$ and $\rho^0$ vector mesons. In addition to that, we present the ratios of the longitudinal to the transverse cross sections that are sensitive to the vector meson wavefunctions. In the calculations of cross sections, Eqs. \eqref{eqn6}, \eqref{eqn7}, \eqref{eqn8}, we need mainly two informations: the scattering amplitude and the overlap wave functions for the vector mesons. For the scattering amplitude, we use the solution of the BK equation presented in the previous section. For the overlap wave functions, we employ two famous models, the Gaus-LC~\cite{r61} and the BG~\cite{r62} models. The effects of these two models on the calculations of the cross-sections for vector mesons are shown here. We compare our theoretical prediction with the existing experimental data for the sake of the validity of our analysis.

\begin{figure}[h]
	\centering
	\begin{subfigure}{0.45\textwidth}
		\centering
		\includegraphics[width=\textwidth]{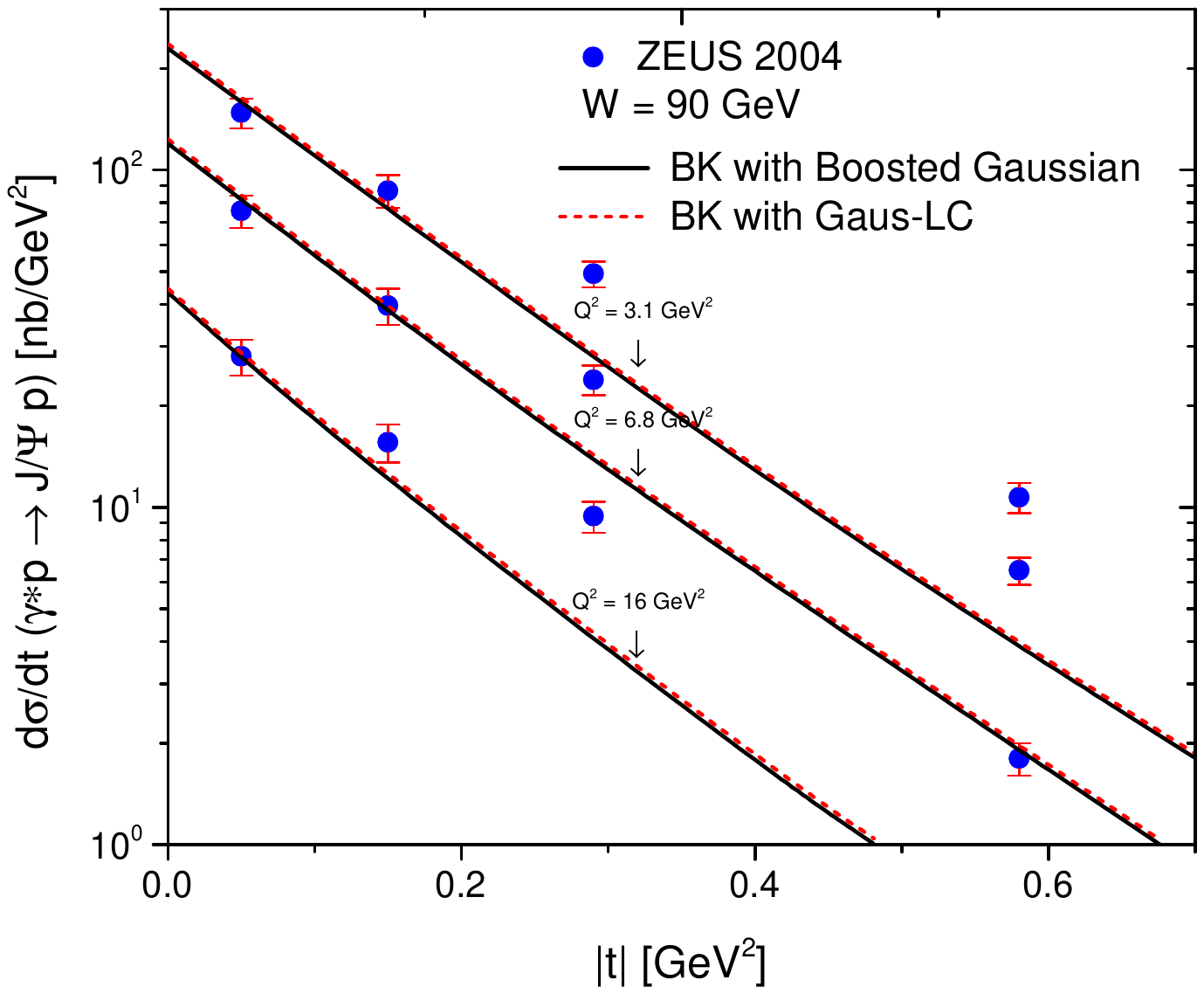}
	\end{subfigure}
	%\hfill
	\begin{subfigure}{0.45\textwidth}
		\centering
		\includegraphics[width=\textwidth]{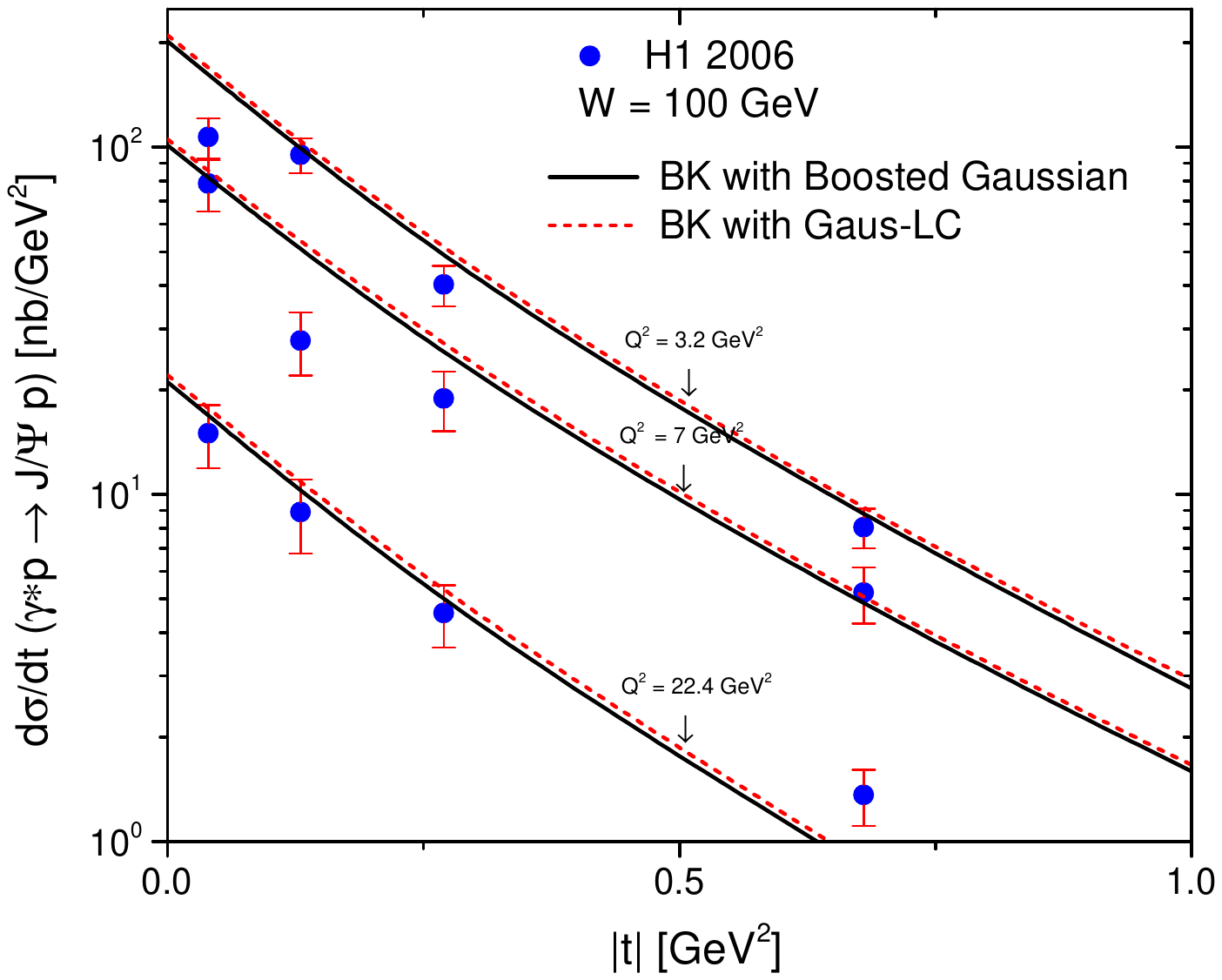}
	\end{subfigure}
	\caption{{\label{fig3}} Differential cross section $d\sigma/dt(\gamma^{\ast}p\rightarrow J/\Psi \hspace{1mm}p)$ vs. $|t|$ calculated using the BK solution with two different vector meson wave functions compared with the experimental data from ZEUS 2004~\cite{r20} at $W=90$ GeV (left) and H1 2006~\cite{r22} at $W=100$ GeV (right).}
\end{figure}

\begin{figure}[h]
	\centering
	\begin{subfigure}{0.45\textwidth}
		\centering
		\includegraphics[width=\textwidth]{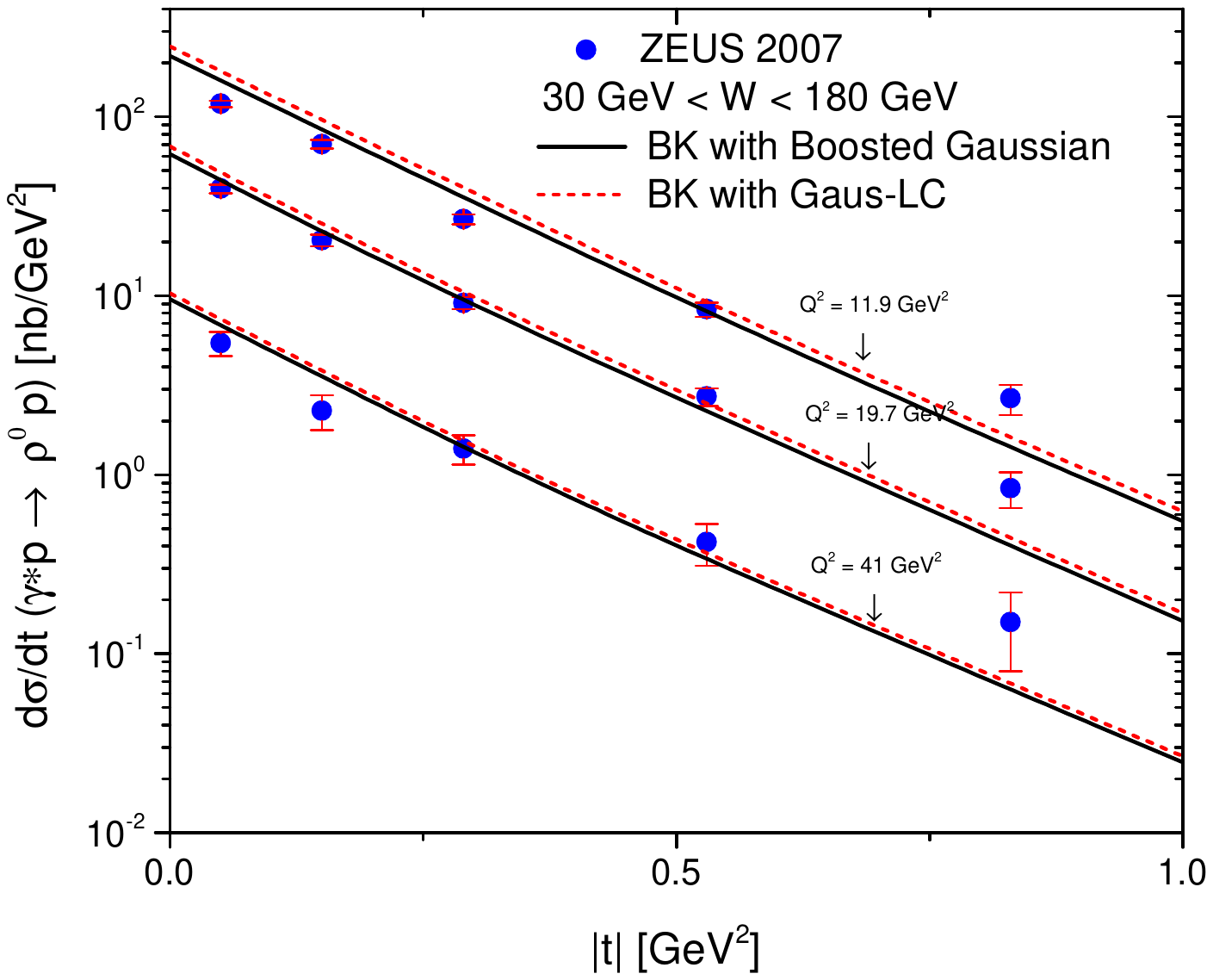}
	\end{subfigure}
	%\hfill
	\begin{subfigure}{0.45\textwidth}
		\centering
		\includegraphics[width=\textwidth]{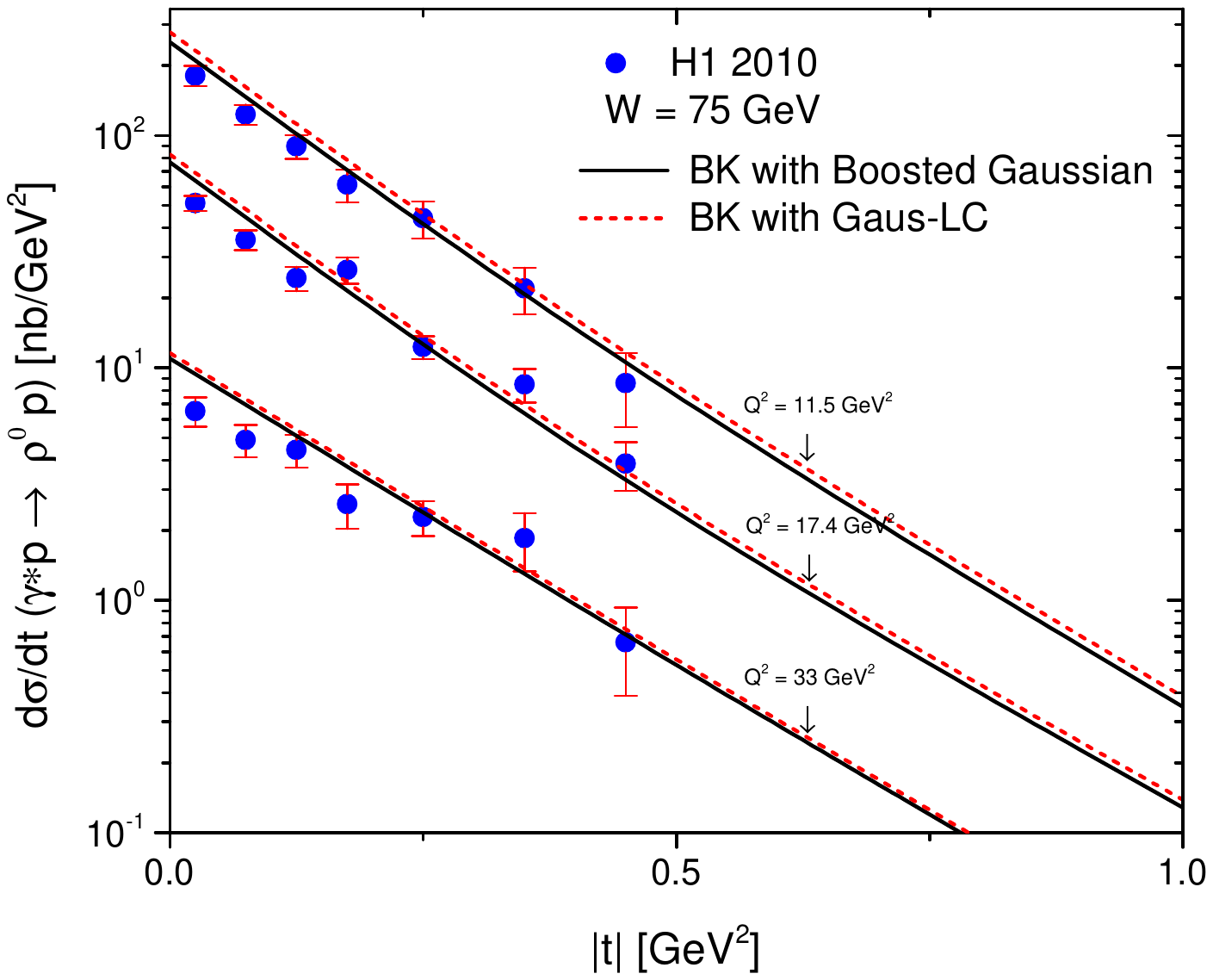}
	\end{subfigure}
	\caption{{\label{fig4}}Differential cross section $d\sigma/dt$ ($\gamma^{\ast}p\rightarrow \rho^0 p$) vs. $|t|$ calculated using the BK solution with two different vector meson wave functions compared with the experimental data from ZEUS 2007~\cite{r23} for $30 \hspace{1mm} \text{GeV}<W<180 \hspace{1mm}\text{GeV}$(left) and H1 2010~\cite{r24} at $W=75$ GeV (right).}
\end{figure}

\begin{figure}[h]
	\centering
	\begin{subfigure}{0.45\textwidth}
		\centering
		\includegraphics[width=\textwidth]{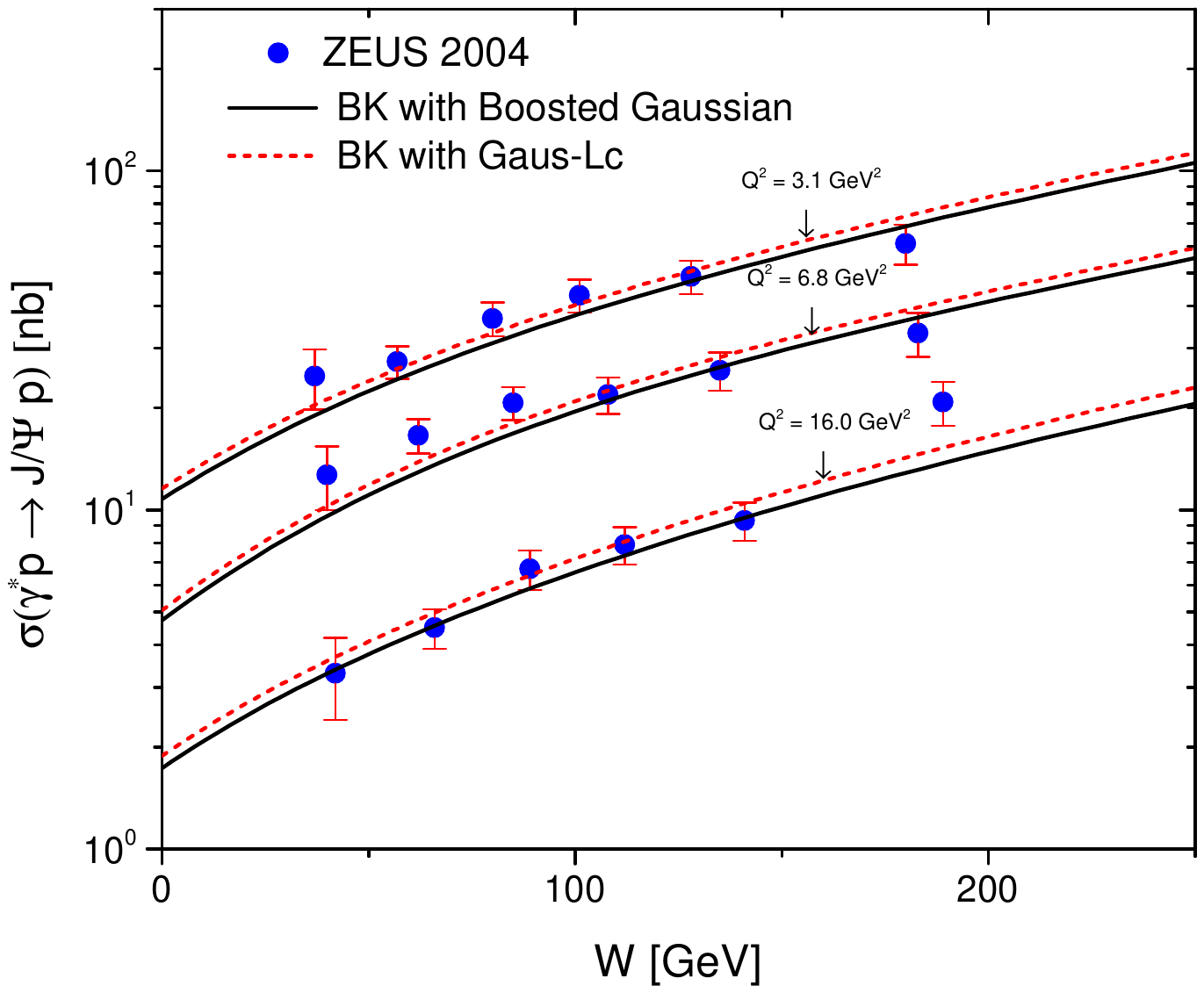}
	\end{subfigure}
	%\hfill
	\begin{subfigure}{0.45\textwidth}
		\centering
		\includegraphics[width=\textwidth]{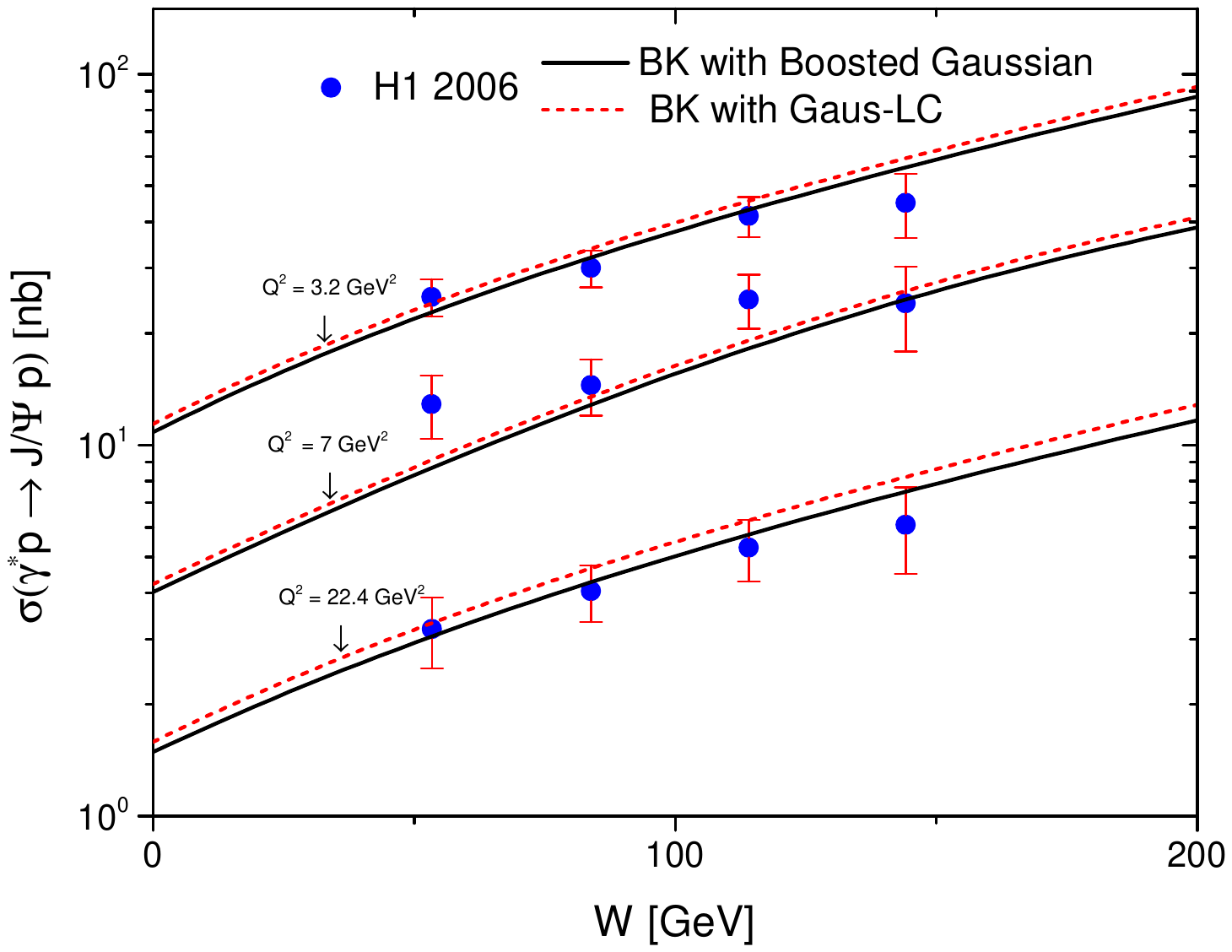}
	\end{subfigure}
	\caption{{\label{fig5}} Total cross section $\sigma (\gamma^{\ast}p\rightarrow J/\Psi \hspace{1mm} p)$ vs. $W$ calculated using the BK solution with two different vector meson wave functions compared with experimental data at various $Q^2$ values from ZEUS 2004~\cite{r20} (left) and H1 2006~\cite{r22} (right).}
\end{figure}

\begin{figure}[h]
	\centering
	\begin{subfigure}{0.45\textwidth}
		\centering
		\includegraphics[width=\textwidth]{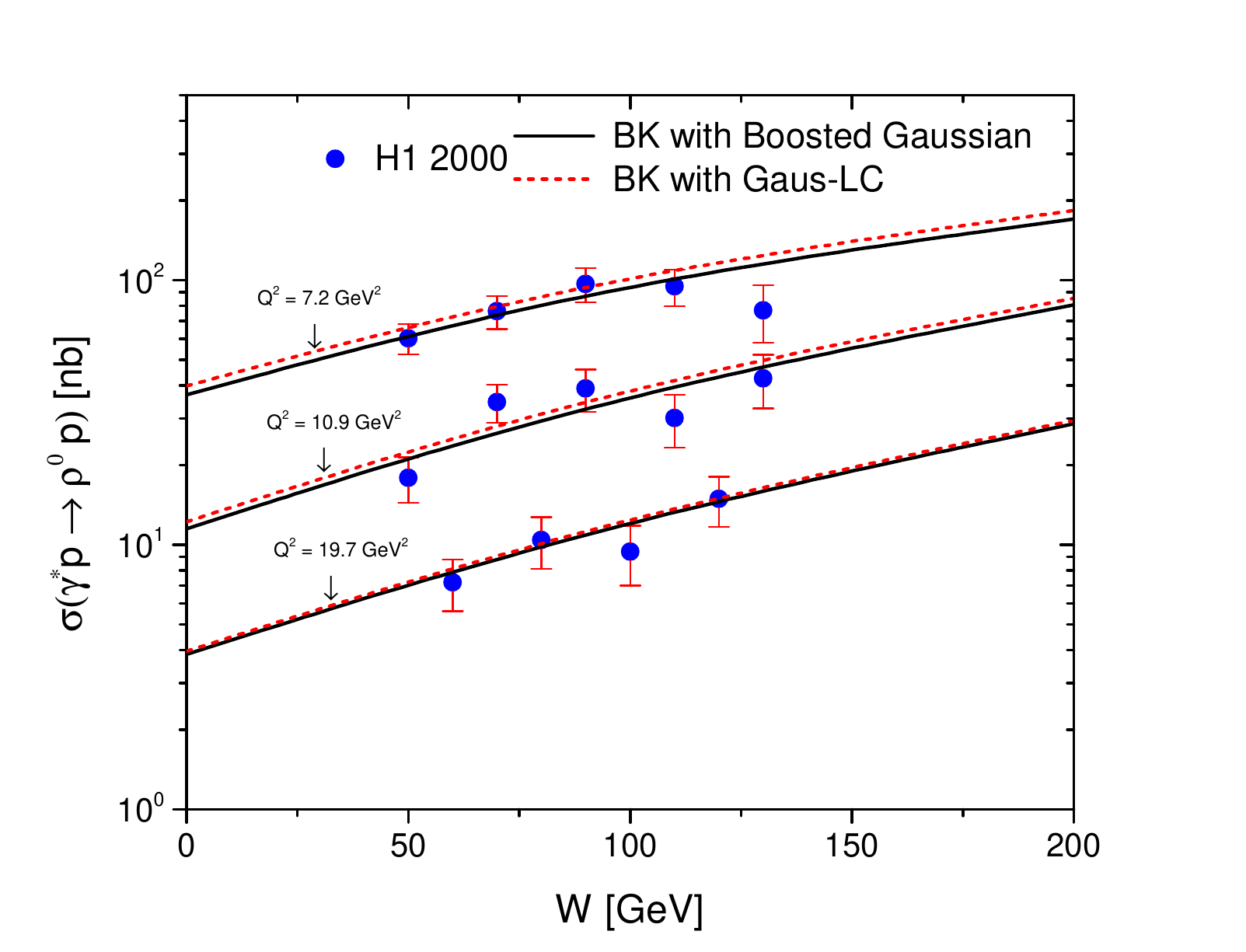}
	\end{subfigure}
	%\hfill
	\begin{subfigure}{0.45\textwidth}
		\centering
		\includegraphics[width=\textwidth]{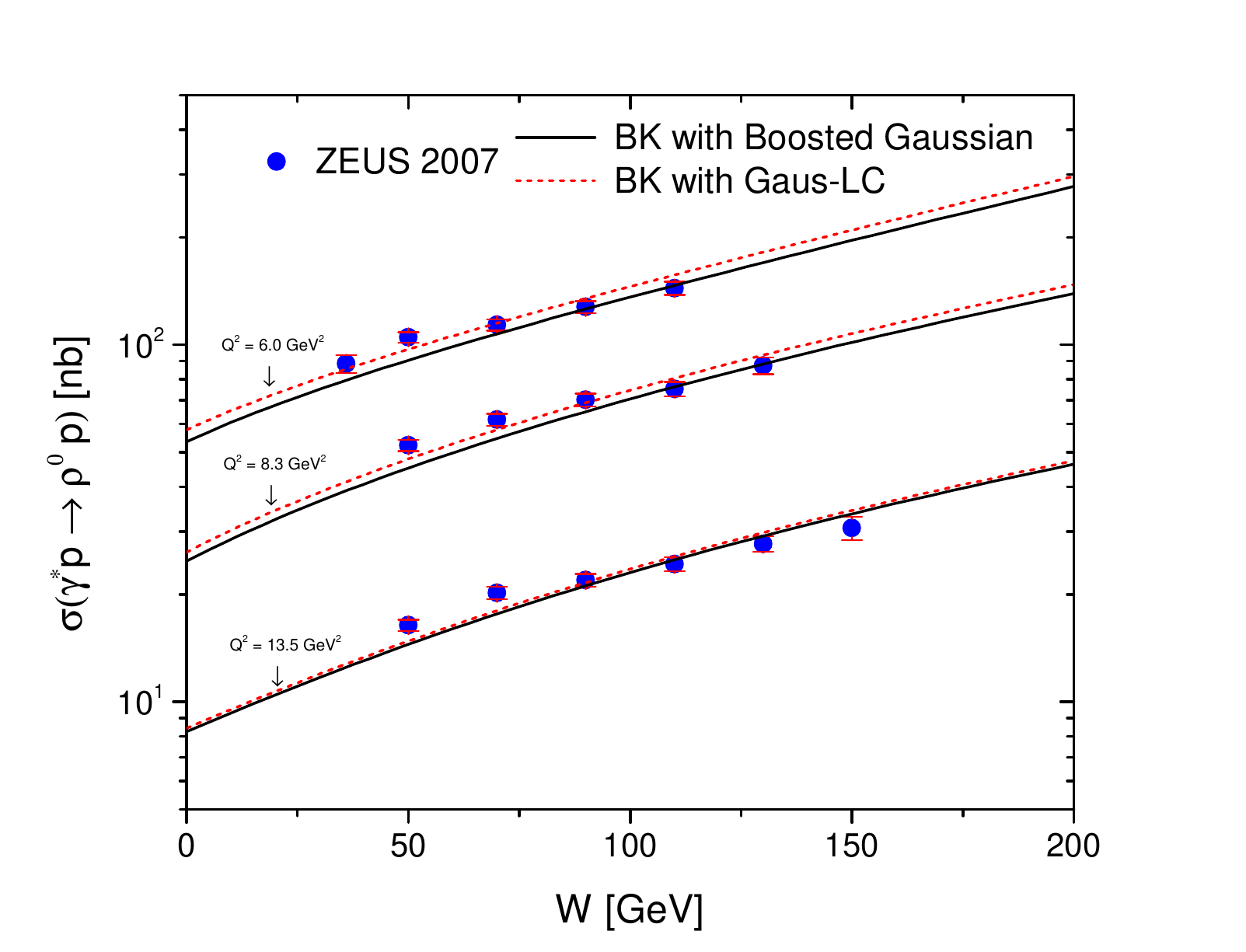}
	\end{subfigure}
	\caption{{\label{fig6}}Total cross section $\sigma (\gamma^{\ast}p\rightarrow \rho^0 p)$ vs. $W$ calculated using the BK solution with two different vector meson wave functions compared with experimental data at various $Q^2$ values from H1 2000~\cite{r19} (left) and ZEUS 2007~\cite{r23} (right).}
\end{figure}

In the first analysis, we present the results of the differential cross sections of the $J/\Psi$ and $\rho^0$ vector meson production as a function of the squared momentum transfer $t$ for various $Q^2$ values of experimtal data. Our results, with the comparison with the experimental data, are shown in Figures \ref{fig3} and \ref{fig4}. The dependence of the total cross sections of the $J/\Psi$ and $\rho^0$ vector mesons on the center-of-mass energy of the $\gamma^{\ast}p$ system $W$, for various $Q^2$ values is presented in Figures \ref{fig5} and \ref{fig6}, respectively. Our calculations of exclusive vector meson productions agree well with the existing experimental data. The results shown here are slightly sensitive to the two vector meson wave function models, the Gaus-LC and the BG models. It is seen from Figure \ref{fig2} that there are differences between these two models that affect the vector meson production calculations. For the transversely polarised $\rho^0$, the Gaus-LC and the BG models are the same, compared to other vector meson parts. Still, they lead to similar predictions of vector meson productions with our BK solution.

\begin{figure}[h]
	\centering
	\begin{subfigure}{0.45\textwidth}
		\centering
		\includegraphics[width=\textwidth]{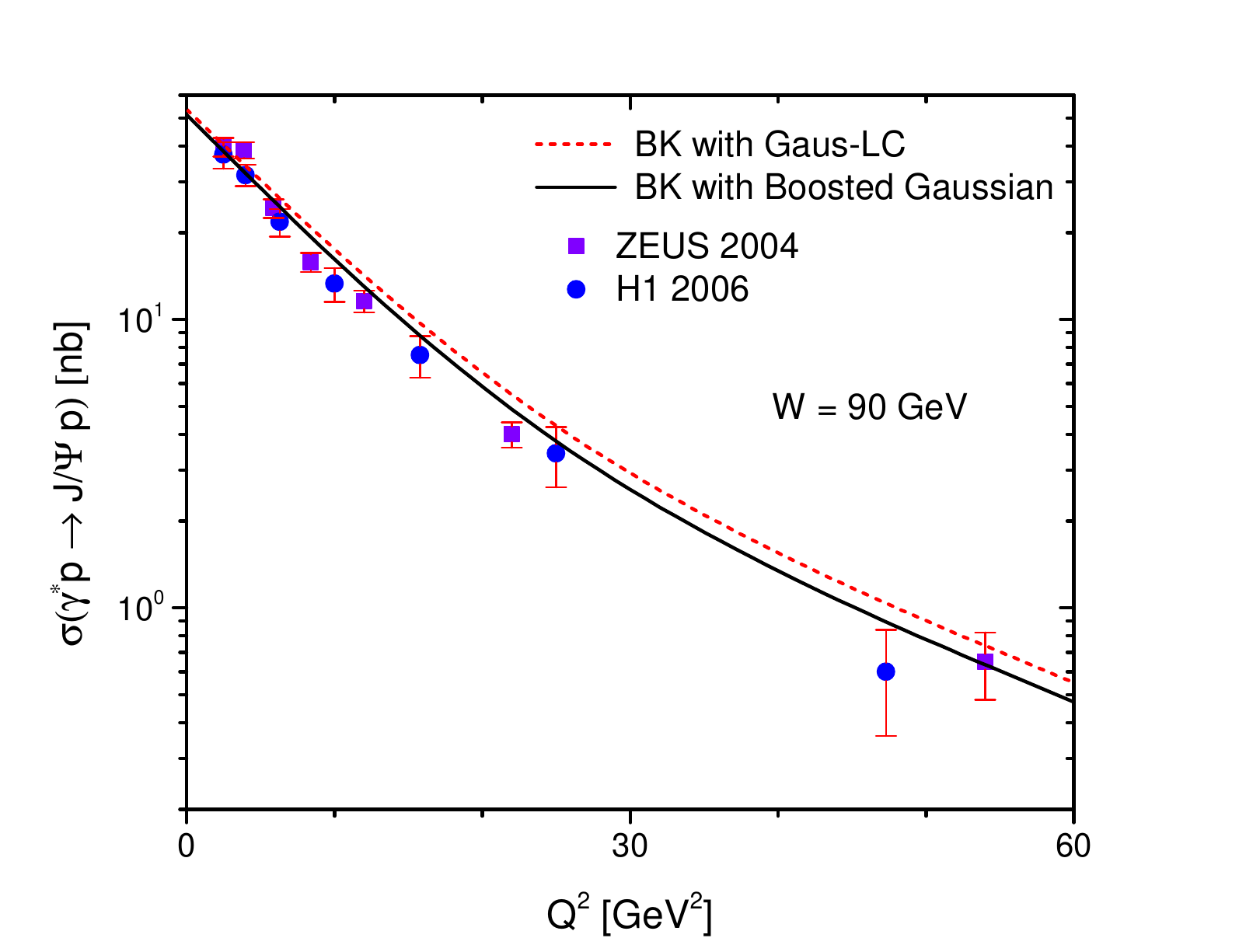}
	\end{subfigure}
	%\hfill
	\begin{subfigure}{0.45\textwidth}
		\centering
		\includegraphics[width=\textwidth]{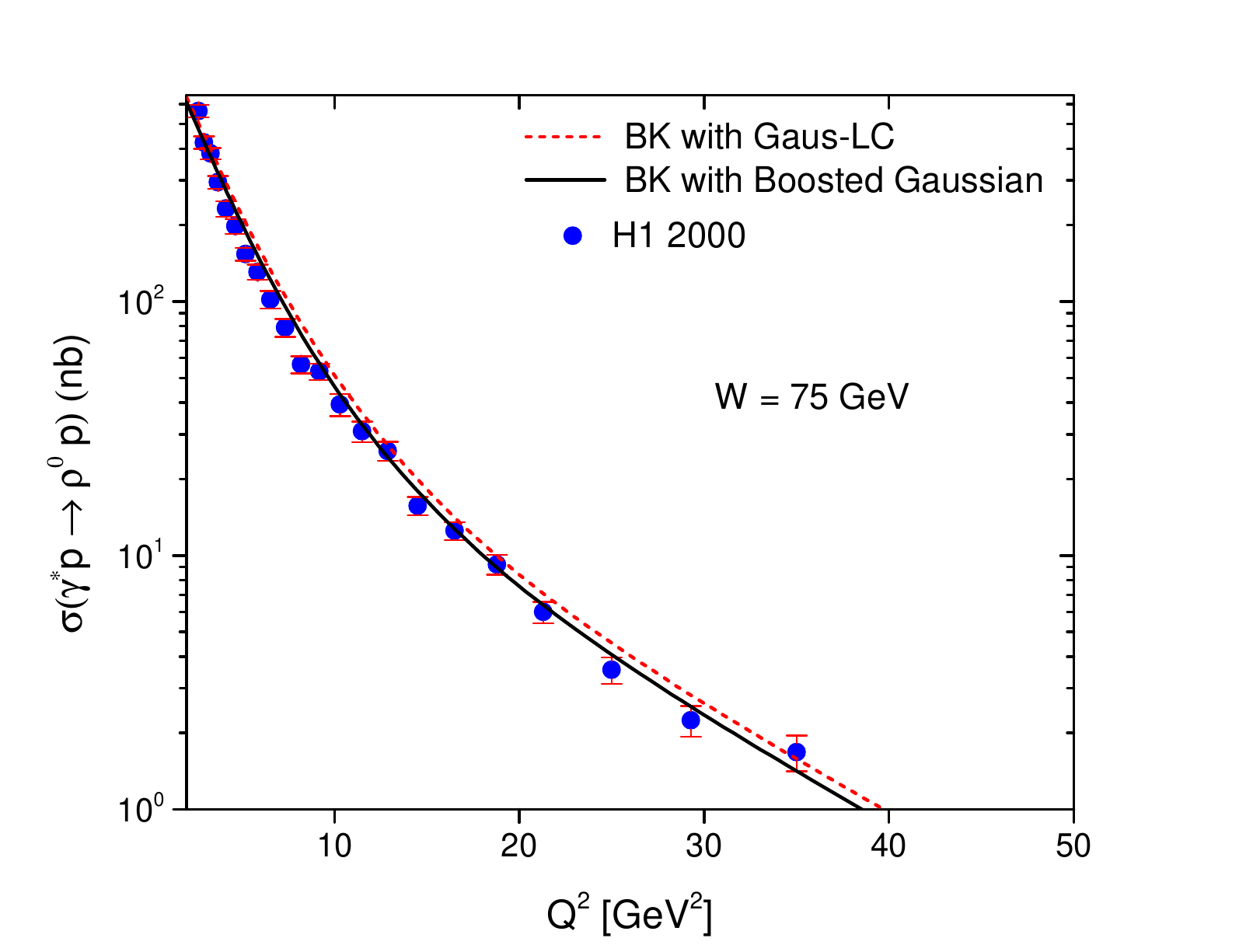}
	\end{subfigure}
	\caption{{\label{fig7}} Total cross section $\sigma$ vs. $Q^2$ calculated using the BK solution with two different vector meson wave functions. (left) For $J/\Psi$ meson at $W=90$ GeV compared with experimental data from ZEUS 2004~\cite{r20} and H1 2006~\cite{r22}. (right) For $\rho^0$ meson at $W=75$ GeV compared with the data from H1 2000~\cite{r19} and ZEUS 2007~\cite{r23}.} 
\end{figure}

In the second analysis, we measure the total cross section, $\sigma$ for the $J/\Psi$ and the $\rho^0$ vector meson production as a function of $Q^2$. In the left plot of Figure \ref{fig7}, we present the total cross section for $J/\Psi$ when $W=90$ GeV, and in the right plot, we present the total cross section for $\rho^0$ when $W=75$ GeV. We can see that the theoretical predictions agree well with the existing experimental data. 

\begin{figure}[h!]
	\centering
	\begin{subfigure}{0.45\textwidth}
		\centering
		\includegraphics[width=\textwidth]{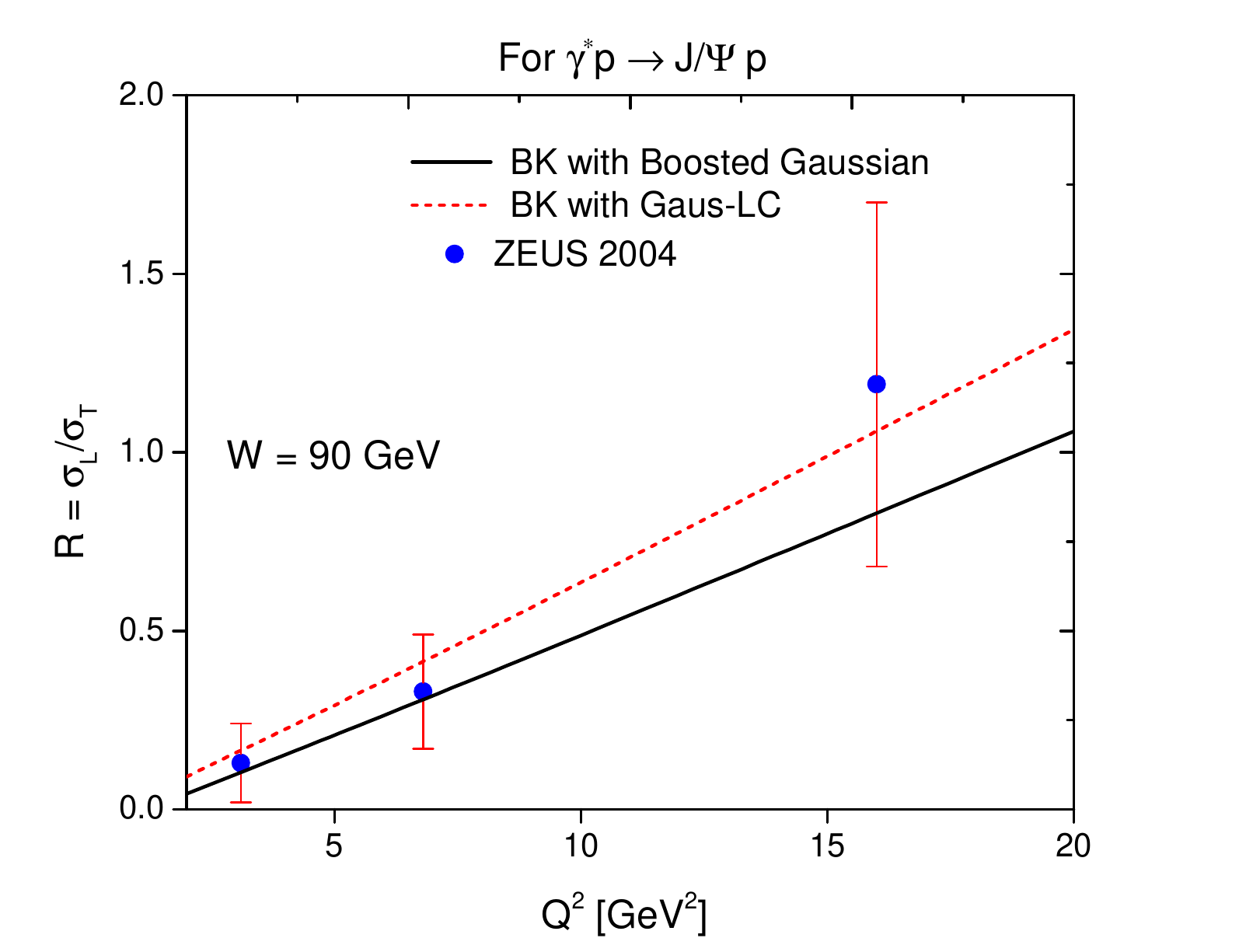}
	\end{subfigure}
	%\hfill
	\begin{subfigure}{0.45\textwidth}
		\centering
		\includegraphics[width=\textwidth]{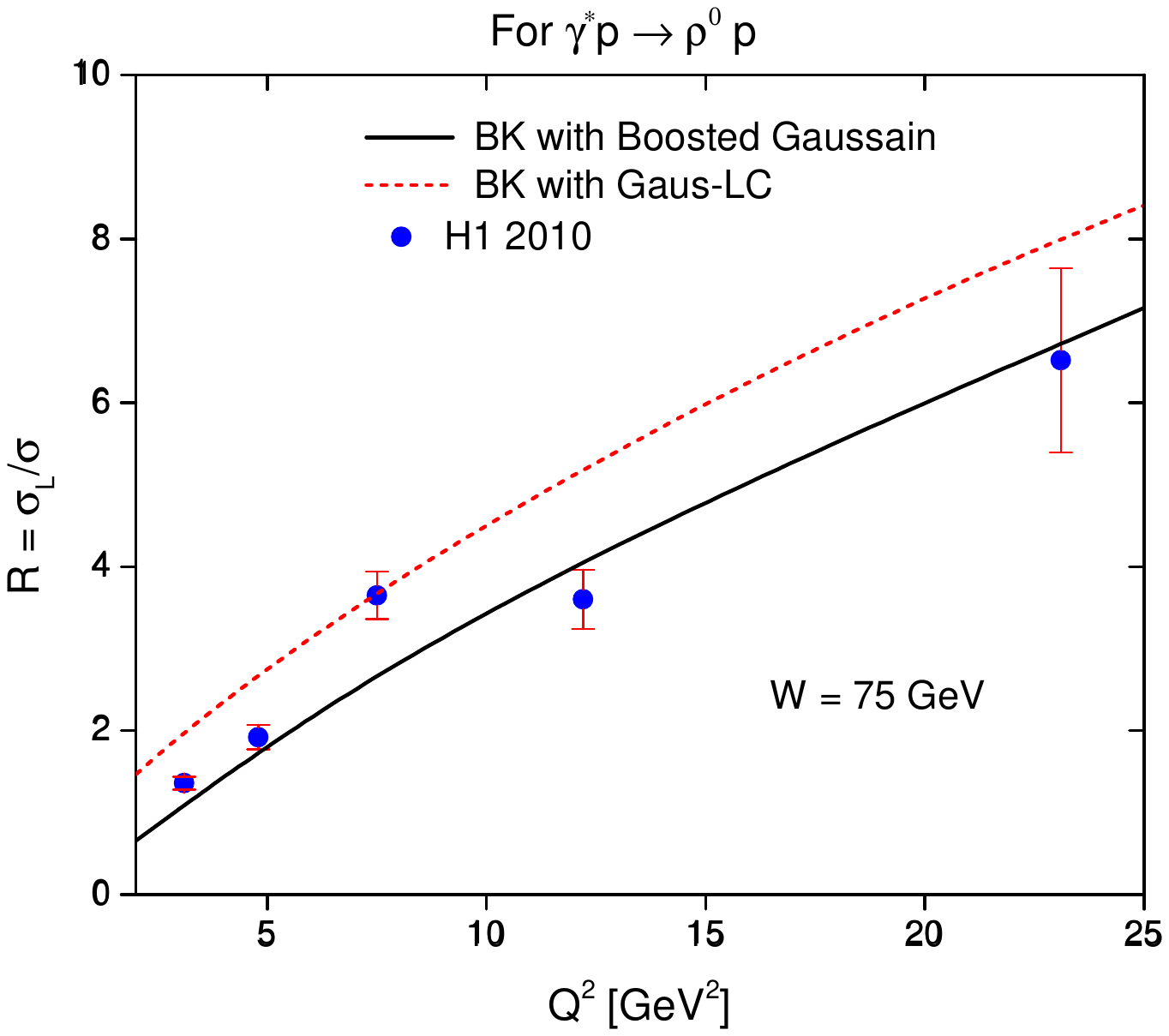}
	\end{subfigure}
	\caption{{\label{fig8}}The ratio $R$ vs. $Q^2$ calculated using the BK solution with two different vector meson wave functions: (left) $R$ for $J/\Psi$ meson at $W=90$ GeV compared with the data from ZEUS 2004~\cite{r20} and (right) $R$ for $\rho^0$ meson at $W=75$ GeV compared with the data from H1 2010~\cite{r24}.}
\end{figure}

Finally, we present the ratio of the longitudinal to the transverse cross sections, $R = \sigma_L/\sigma_T$, as a function of $Q^2$ at fixed $W$. It is seen from Figure \ref{fig8} that the ratio for the $J/\Psi$ vector meson is proportional to the $Q^2$, and for the $\rho^0$ vector meson, the ratio grows rapidly as $Q^2$ increases. For both vector mesons, as $Q^2$ increases, we see vast differences between the two vector meson wave function models. Thus, the ratio is highly sensitive to the given vector meson wave function models at high $Q^2$.
\section{Summary and conclusion}\label{sec5}
In this paper, we investigated exclusive vector meson production within the color dipole description of DIS. Using an analytical solution of the BK equation or the BK evolution theory, we calculated cross-sections of the $J/\Psi$ and $\rho^0$ vector mesons as functions of $t$, $W$, and $Q^2$. To validate our analysis, we compared the theoretical predictions for cross-sections of $J/\Psi$ and $\rho^0$ vector mesons with existing experimental data, finding reasonable agreement between them. Our investigation began with a brief overview of the color dipole description of DIS for vector meson production. Within this framework, we provided the scattering amplitude for vector meson production. This amplitude required two essential components for vector meson production: the vector meson wavefunctions and the dipole-proton scattering amplitude. Employing two well-known models, the Gaus-LC and the BG, for the vector meson wave functions, and obtaining the dipole-proton scattering amplitude from the solution of the BK equation, we computed the cross-sections of $J/\Psi$ and $\rho^0$ vector mesons, along with the ratios of longitudinal to transverse cross-sections.

The results demonstrate that our analytical solution to the BK equation effectively computes vector meson production within a low $Q^2$ range. The BK equation performs well in low $Q^2$ and small-$x$ domains. However, for high $Q^2$, corrections from the DGLAP evolution equation are necessary and should be incorporated into the calculations. Nonetheless, our analytical solution remains valid for measuring vector meson cross-sections within the specified range. Moreover, our results highlight the sensitivity of outcomes to the Gaus-LC and BG models. While differences between these models are less noticeable for the $\rho^0$ meson as $Q^2$ increases, distinctions are evident for the $J/\Psi$ meson even at high $Q^2$. Additionally, it is observed that the longitudinal to transverse cross-section ratios for $J/\Psi$ and $\rho^0$ vector mesons reveal a significant dependency on the two wave function models as $Q^2$ increases. Based on our analyses, we conclude that the results obtained from the BG with BK solution provide a better description of the data compared to those derived from the Gaus-LC with BK solution. 

Exclusive vector meson production is an excellent method to probe the properties of nuclear matter. Our analytical solution to the BK equation has the ability to address further phenomenological studies at both existing and future experimental facilities. We anticipate that diffractive processes such as exclusive vector meson production at forthcoming experimental facilities will shed light on numerous aspects of QCD.

\begin{center}
\rule{8cm}{0.3mm}
\end{center}

\end{document}